\documentstyle[a4,psfig]{article}
\begin{document}
\def \beq{\begin{equation}}
\def \eeq{\end{equation}}
\def \beqar{\begin{eqnarray}}
\def \eeqar{\end{eqnarray}}
\def \g{{\rm GeV}}
\vspace{0.5in}
\centerline{\bf Stark quenching for the $1s^22s2p^3P_0$ 
level in beryllium-like ions }
\centerline{\bf and parity violating effects}
\vspace{0.5in}
\centerline{\it M.~Maul, A.~Sch\"afer}
\centerline{\it Naturwissenschaftliche Fakult\"at II}
\centerline{\it  -- Physik --}
\centerline{\it der Universit\"at Regensburg}
\centerline{\it 93040 Regensburg}
\bigskip
\centerline{and}
\bigskip
\centerline{\it P.~Indelicato}
\centerline{\it Laboratoire de Physique Atomique et Nucl\'eaire, Bo\^ite 93,}
\centerline{\it Universit\'e Pierre et Marie Curie, 4 place Jussieu,}
\centerline{\it F--75252 Paris CEDEX 05, France}
\date{ }

\bigskip

\centerline{\bf ABSTRACT}
\medskip
\begin{quote}
In this paper we present some concepts in 
heavy ion atomic physics for the extraction of 
parity violating effects. 
We investigate the effects of the
so-called Stark-quenching, i.e., the fast decay
of a meta stable state induced by a Stark field,  and the
superposition of one- and two-photon transitions
in beryllium-like heavy ions. 
It turns out that the discussed theoretical phenomena for heavy ions
with few electrons are beyond the scope of present day experimental
possibilities because one has to require 
beam energies of up to 1 TeV/A, laser intensities of
 up to $10^{17}  {\rm W/cm^2}$ and ion currents of up to 
$10^{11}$ ions per second in beryllium-like uranium.  
However,  especially the
superposition of one- and two-photon transitions is a
very interesting phenomenon that could provide the germ
of an idea to be applied in a more favorable system. 
\end{quote}
PACS: 34.50 Rk, 34.80 Qb
\newpage
\noindent
\section{Introduction}
The experiments done recently at CERN allow  a very precise and 
excellent determination of the constants of the standard model
of weak interactions \cite{Al96}. However, it is important to test
the theory also for small momentum transfers, where possibly the
situation could be essentially different from LEP experiments.
Such small momentum transfer experiments have been performed 
in several heavy atoms \cite{Ro95}. Recently the atomic parity
violating experiment with cesium has been improved in precision
for a factor of seven, so that one obtains a signal even for
the anapole moment \cite{Le97}, which underscores the great potential
which still lies in the atomic physics to compete with the high energy
experiments for measurements of weak constants and phenomena. 
\newline
\newline
In principle heavy ions offer an even better access to weak interaction
processes than heavy atoms because of the large overlap between the
nucleus and the electrons in inner shells. 
%
%
On the other hand this large overlap with the nucleus inserts the
influence of nuclear effects. Recently, those effects have been 
studied intensively
in hydrogen-like heavy ions. The nuclear polarization including
vacuum polarization-nuclear polarization corrections results in an
energy shift in hydrogen-like $^{208}_{82}$Pb for the ground state and
the two first excited levels which is in the range of meV 
\cite{LNPBG96}. In the same range is the nuclear recoil effect in
hydrogen-like uranium \cite{ASY95}. Another effect comes from the
uncertainty of the nuclear radius which causes for the 1S energy of 
hydrogen-like uranium an uncertainty of 0.1 eV \cite{BMPPGS97}. To give an idea
as to the impact of those nuclear effects, we consider the level difference
between the ground state and the first excited state in beryllium-like
uranium which is approximately 260 eV (c.~f. Tab.~\ref{tab1}). So the nuclear
structure results in a per mille effect which has to be taken into 
account in
all possible analyses of parity violating effects in heavy ions with few
inner shell electrons. 
Detailed analyses
for helium-like uranium \cite{Schaefer} and uranium with up to five
electrons \cite{Maul}  were undertaken to find a realistic experimental scheme to extract the
pv signal. In this contribution we continue our search by analyzing a 
number of possible signals based on the inclusion of the Stark
effect for analyzing parity violation effects
in heavy ions.
\newline
\newline
The first excited state of beryllium-like uranium $1s^22s2p^{\;3}P_0$, 
in the case of zero nuclear spin, is meta stable because it can decay
only by a two-photon transition to the ${1s^22s^2} ^{\;1}S_0$ ground state.
In case of a non-vanishing nuclear spin the $1s^22s2p^{\;3}P_0$ state 
gets an admixture from a $ |^{\;3}P_1\rangle $ state lying closely above 
due to hyperfine mixing which drastically reduces the lifetime 
of the meta stable
$1s^22s2p^{\;3}P_0$ state \cite{MPI93}. 
%
%
While this 'Hyperfine Quenching'  is
due to a magnetic field, in this contribution we want
to investigate a similar effect 
due to an electric field, namely the 'Stark Quenching'
in beryllium-like heavy ions.
The principle to use the Stark effect for measuring atomic parity
violation can be found already in \cite{BB75,LW75} and has been 
exploited extensively in earlier parity violating experiments in 
heavy atoms \cite{BP83,NMW88}.
The strength of the Stark effect which 
leads to a mixing of the levels $|^{\;1}S_0\rangle$ and $ |^{\;3}P_1\rangle $
can be varied by experimental conditions. 
This is interesting for two reasons. 
%
%
On the one hand it provides an alternative means for measuring the
the parity violating mixing of the
two levels $1s^22s2p^{\;3}P_0$ and ${1s^22s^2} ^{\;1}S_0$
by adding another mixing between the 
$1s^22s2p^3P_1$ and the ${1s^22s^2}^1S_0$ states. 
On the other hand 
it allows for a control  of the lifetime of an atomic state
by experimental conditions, which will certainly be useful for other
atomic physics experiments.
\section{Parity admixture between the $1s^22s2p^{\;3}P_0$ and  
${1s^22s^2}^{\;1}S_0$
         in beryllium-like heavy ions}
\begin{figure}
\psfig{figure=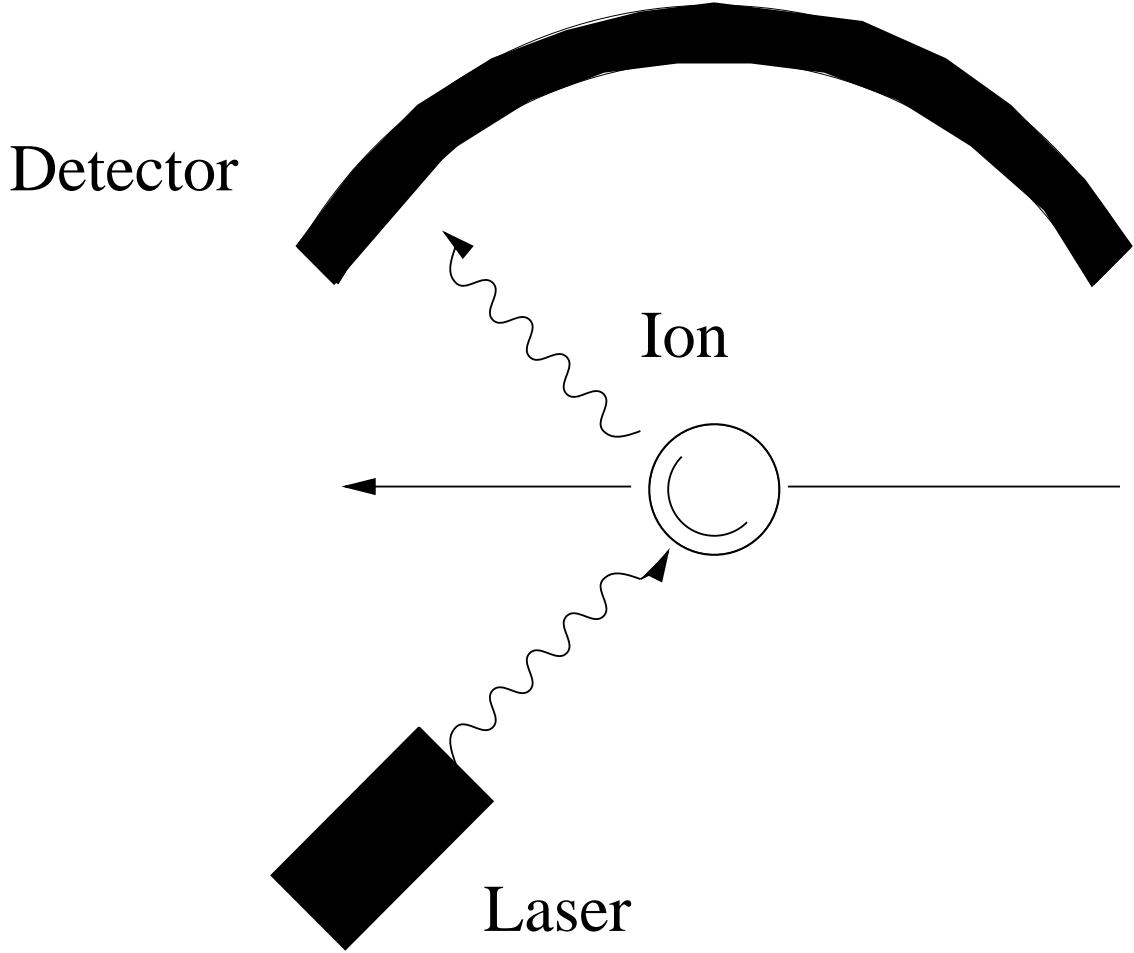,height=5in}
\caption{Experiment for a direct excitation of the
parity violating 2-photon transition.}
\label{fig1}   
\end{figure}

The first five atomic states of the beryllium-like heavy ions are 
\beqar
\label{2}  
|0\rangle &=& |^{\;1}S_0 \rangle = | 1s^2 2s^2        \; J=0 \rangle
\nonumber \\
|1\rangle &=& |^{\;3}P_0 \rangle = | 1s^2 2s 2p_{1/2} \; J=0 \rangle
\nonumber \\
|2\rangle &=& |^{\;3}P_1 \rangle =\alpha | 1s^2 2s 2p_{1/2} \; J=1 \rangle
                                 +\beta  | 1s^2 2s 2p_{3/2} \; J=1 \rangle
\nonumber \\
|3\rangle &=& |^{\;1}P_1 \rangle =\beta  | 1s^2 2s 2p_{1/2} \; J=1 \rangle
                                 -\alpha | 1s^2 2s 2p_{3/2} \; J=1 \rangle    
\nonumber \\
|4\rangle &=& |^{\;3}P_2 \rangle = | 1s^2 2s 2p_{3/2} \; J=2 \rangle
\quad.
\eeqar
From the  interaction due to the exchange of neutral
$Z^0$ bosons between nucleus and electron shell one
derives the Hamiltonian
\beq
\label{3}
H_{\rm pv} = \frac{G_F}{2 \sqrt{2}}
             (1 - 4 \sin^2 \vartheta_{\rm W} - \frac{N}{Z})
                              \rho  \gamma_5 
\quad.
\eeq
$G_F$ denotes Fermi's constant, $\vartheta_W$ the Weinberg angle, $N$
the neutron number, $Z$ the proton number, and $\rho$ the nuclear density
normalized to $Z$. This formula also demonstrates why highly charged heavy 
ions with few electrons are proper candidates for investigating
parity non-conservation effects: The wave function admixture coefficient 
$\eta_{\rm pv}$ which is given by
\beq
\label{5}
\eta_{\rm pv} = \frac{\langle i | \frac{G_F}{2 \sqrt{2}}
             ( 1- 4 \sin^2 \vartheta_{\rm W} - \frac{N}{Z} )
                              \rho \gamma_5   | f \rangle}{E_i - E_f}
= \frac{\langle i |H_{\rm pv} |f \rangle}{E_i - E_f}
\eeq
is very large (typically orders of magnitude larger than for
the outer shell in neutral atoms) due to the big overlap
between the nucleus and the electron states. The admixture is then 
described by the following matrix

\beq
\label{4}
H_{\rm tot} = \left[ \begin{array}{cc}
           E_0 + \frac{i}{2} \Gamma_0 & W(0,1)                     \\
           W(1,0)                     & E_1 + \frac{i}{2} \Gamma_1 \\ 
               \end{array} \right] \quad,
\eeq
where $W(0,1) = \langle 0 | H_{\rm pv} |1 \rangle$.  
The expression $\eta_{\rm pv}$ is modified by radiative corrections 
\cite{MS83}, which together with other electroweak precision experiments
give  valuable constraints as to the mass of the Higgs boson and
possible other particles connected to new physics \cite{Ro95,Ro97,De97}.
\newline
\newline
Tab.~\ref{tab1} shows the various parity mixing coefficients $\eta_{\rm pv}$ 
for some stable beryllium-like heavy ions from $Z=26$ to $Z=92$ with zero
nuclear spin. In this table there is given also the energies of the 
two levels $1s^22s2p^{\;3}P_0$ and ${1s^22s^2} ^{\;1}S_0$. 
The parity admixing effect is decreasing by four
orders of magnitude from uranium to iron. 
How can one measure those admixture coefficients? There is in principle
the possibility to use a laser to excite the $1s^22s2p^{\;3}P_0$ state starting
from the ground state. One may use an ordinary optical
laser and tune the  energy with the help of the relativistic Doppler shift
by choosing a particular angle towards the ion beam. 
As the laser light is coherent, in principle, it should only
excite the parity violating 2E1 transition (and to a lower extent
also 2M1). Therefore the transition amplitude will be proportional
to $\eta^2_{\rm pv}$ which is an extremely small number. 
For the  energy gap of 260 eV to be covered  
for the beryllium-like uranium ion by
an 1 eV laser by means of the Doppler shift,
 this would require ion energies  of
260  GeV/u. This could be well covered by 
the planed LHC accelerator which is supposed to reach 2.76 TeV/u 
in Pb onto Pb collisions.
For lighter ions this figure reduces accordingly.
The experimental setup is sketched in Fig. \ref{fig1}.
\newline
\newline
In order to estimate  the expected transition rates, we must
determine at least approximately the two-photon transition amplitude. 
In principle one has to 
sum over all intermediate states. 
%
%
Here we only want to give a rough estimate taking into consideration 
only the states $|^{\;1}S_0\rangle$,
$|^{\;3}P_0\rangle$, and $|^{\;3}P_1\rangle$.
Furthermore we  use the one-photon transition matrix elements, to estimate
the two-photon induced amplitude. To be more precise we
use the following formulas for the spontaneous and induced transition
rates (in atomic units). 
For the laser intensity distribution we take
a simple box form with full width $\Gamma_{\rm laser}$ and intensity
$I_0$. In case of the two-photon transition the laser energy is chosen to be
half of the transition energy:
$ \omega_{\rm laser} = \omega_{i \to f} /2 $. In atomic units we obtain:
\begin{eqnarray}
\label{5.1}
W_{i \to f} ({\rm 1 \;Photon\; spontaneous})
   &=& \sum_{i,f,M,\lambda} \frac{1}{2j_i+1} \frac{\omega_{i \to f}}{2\pi c} 
      |\langle f | a_{1,M}^{(\lambda)} | i \rangle |^2
\nonumber \\
\nonumber \\
W_{i \to f} ({\rm 1 \;Photon\; induced})
   &=& \frac{(2 \pi)^2  c I_0}{\omega_{i \to f}^2 \Gamma_{\rm laser}}
      \frac{1}{2j_i+1} 
      |\sum_{i,f,M}\vec \epsilon \cdot \vec Y^{(\lambda)}_{1M}(\hat {\vec k})
                    \langle f |  
                  a_{1,M}^{(\lambda)} | i \rangle |^2
\nonumber \\
\nonumber \\
W_{i \to f} ({\rm 2 \;Photon\; induced})
&=& \frac{(2 \pi)^3 c^2 I_0^2}{\omega_{\rm laser}^4 \Gamma_{\rm laser}}
 \frac{1}{2j_i+1} 
\nonumber 
\end{eqnarray}
\begin{equation}
\times\left|\sum_{i,f,M_1,M_2,n,\lambda_1,\lambda_2} 
       \vec  \epsilon_1 \cdot  \vec Y^{(\lambda_1)}_{1M_1}( \hat {\vec k}) 
       \vec  \epsilon_2 \cdot  \vec Y^{(\lambda_2)}_{1M_2}( \hat {\vec k})
       \left\{\frac{\langle f |a_{1,M_1}^{(\lambda_1)}|n \rangle
                    \langle n |a_{1,M_2}^{(\lambda_2)}|i \rangle}
                   { E_i - E_n - \omega_{\rm laser}} + ( 1 \leftrightarrow 2)
       \right\} \right|^2 \quad.
\end{equation}
$\vec Y^{(\lambda)}_{LM}(\hat {\vec k}) a_{L,M}^{(\lambda)}$
are the usual terms of the multipole expansion into electric
$(\lambda = 1)$ and magnetic $(\lambda = 0)$ $2^L$-pole components.
$\vec \epsilon$ denotes the photon polarization vector. 
The sum $\sum_{i,f,\dots}$ is the sum over the $m$-quantum numbers of
the initial state $i$, the final state $f$ etc.
In case
of the two-photon transition we must distinguish between the
polarization of the first $\vec \epsilon_1$ and the second
 $\vec \epsilon_2$ photon.
$\hat {\vec k}$ is the unit vector of the photon momentum.
This gives for the two spontaneous transitions
between the $|^{\;1}S_0\rangle$, $|^{\;3}P_0\rangle$, 
and $|^{\;3}P_1\rangle$ states: 
(Note that in case of the two-photon
transition $\hat {\vec k}$ is the same for both photons.)

\begin{eqnarray}
W_{\rm 3P1 \to 1S0} (E1) &=& 
\frac{1}{3}
\frac{\omega_{\rm 3P1 \to 1S0}}{2 \pi c}
| \langle ^1S_0 || E1 || ^3P_1 \rangle |^2
\nonumber \\
W_{\rm 3P1 \to 3P0} (M1) &=&
\frac{1}{3}
\frac{\omega_{\rm 3P1 \to 3P0}}{2 \pi c}
| \langle ^3P_0 || M1 || ^3P_1 \rangle |^2 \quad.
\end{eqnarray}
With the transition probabilities we get also the reduced matrix elements
which we can in turn use to calculate the transition rates
for induced absorption which is the same for induced emission (without the
contribution from spontaneous emission). 
The $\vec k$ vector of the laser light points
into the x-direction with polarization vector 
$\vec \epsilon = (0,\epsilon_y,\epsilon_z); 
\quad \epsilon_y^2 + \epsilon_z^2=1$. The laser intensity is
given by $I_0$ and the width of the laser is $\Gamma_{\rm laser}$ assuming
a box like frequency distribution. We obtain for
the one-photon and two-photon induced absorption/emission rates:
\begin{eqnarray}
W_{\rm laser;\; pv} (2E1) &=& \frac{1}{2}\eta_{\rm pv}^2 
                 \frac{ \pi c^2 I_0^2}{\Gamma_{\rm laser} 
                 \omega_{\rm laser}^4}
                 \frac{| \langle^1S_0  || E1 ||^3P_1 \rangle |^4}
                      { (E_{\rm 1S0} - E_{\rm 3P1} - \omega_{\rm laser})^2}
\nonumber \\
\nonumber \\
&& \times
\left|  \epsilon_{1z}\epsilon_{2z} 
     +  \epsilon_{1y}\epsilon_{2y} \right|^2
\nonumber \\
\nonumber \\
W_{\rm laser;\; nat} (E1M1) &=& \frac{1}{2}
                  \frac{\pi c^2 I_0^2}{\Gamma_{\rm laser} \omega_{laser}^4}
                 \frac{| \langle ^1S_0 || E1 || ^3P_1 \rangle |^2
                       | \langle ^3P_1 || M1 || ^3P_0 \rangle |^2}
                      {( E_{\rm 1S0} - E_{\rm 3P1} - \omega_{\rm laser})^2}
\nonumber \\
\nonumber \\
&& \times
| \epsilon_{1z}\epsilon_{2y} + \epsilon_{1y}\epsilon_{2z}|^2 \quad.
\end{eqnarray}
%
%
%
%
%
%
In case of the 2E1 transition coherence of the two photons
results in $| \epsilon_{1z}\epsilon_{2z} + \epsilon_{1y}\epsilon_{2y}|^2 = 1$.
For the E1M1 transition we average over all
polarizations of the two photons which gives an additional
factor 1/2. The energy width of the laser has been assumed to be 
$\Gamma_{\rm laser} = 1$ ${\rm  eV}$ in the moving frame. 
We have chosen this 
value because high laser intensities are incompatible with
small line widths.
From Tab.~\ref{tab2} it can be read off that in the case of uranium
for a counting
rate of 1000 photons per second, an apparatus of three meters length
along the beam, and a luminosity of $10^{11}$ ions per second for
inducing the parity conserving amplitude by induced emission 
a laser intensity
of at least $10^{14} W/({\rm cm})^2$ is necessary, which may be 
practically attainable. For the parity violating
amplitude $10^{20} W/({\rm cm})^2$ is required, which 
exceeds present technical possibilities.
%
%
Moreover, for a separation of the parity non-conserving E1E1 transition 
from the unwanted E1M1 transition the latter one has to be suppressed
by means of the laser polarization for at least 12 orders of magnitude in
uranium which does not seem to be a practical scheme either.
The E1M1 amplitude is slightly increasing for lighter ions, but
because of the smaller admixture coefficients $\eta_{\rm pv}$ the
interesting 2E1 transition rate decreases about five orders of magnitude
from uranium to iron.

\section{Stark-quenched amplitude}
It may very well be that the transition amplitudes are too small to
be measured. Therefore it is interesting to search for possible
amplification factors. The idea is analogous to the hyperfine 
quenching technique, but this time it would be done
with a Stark field. To observe such a  quenching effect would
be of great interest in its own. 
We calculate the Stark-quenched amplitude simply by diagonalizing the
matrix
\begin{equation}
\label{6}
H_{\rm tot} = \left[ \begin{array}{cc}
           E_0 + \frac{i}{2} \Gamma_0 & W_{Stark}(0,2)                     \\
           W_{Stark}(2,0)                     & E_2 + \frac{i}{2} \Gamma_2 \\
               \end{array} \right] \quad,
\end{equation}
with 
\begin{equation}
\label{7}
W_{Stark}(i,j) =E_z \langle i| ez | j \rangle
\end{equation}
Here $\vec E = (0,0,E_z)$ is the electric Stark field.
The real part of the 2x2 matrix in Eq.~(\ref{7}) of 
each eigenvalue is the energy of the corresponding level and
the imaginary part is its lifetime. 
In Tab.~\ref{tab3} we give the
resulting lifetime $\tau_{ \rm Stark}$ together with the values of the
energy separation $\Delta E_0$ of the unperturbed states and 
the additional energy shift due to the influence of the Stark field
$\Delta E_{ \rm Stark}$. 
The crucial parameter is obviously the electric field strength which
can be applied to induce the Stark mixing. Present technique allows
up to $E = 3 * 10^8$ V/m.
This would correspond for uranium to a lifetime
of $1.104\times 10^5$ seconds which is still by far too large. 
Through Lorentz contraction the applied electric field for 
an 1 TeV/A - accelerator is increased 
by a factor 1000. This is partly off set by the time dilation of the decay. 
Thus the net gain is $\gamma$ instead of $\gamma^2$. As a result 
the lifetime would be reduced to 110.4 s.

\section{Separating parity violating and parity conserving
         transitions by Stark effect}
\begin{figure}
\psfig{figure=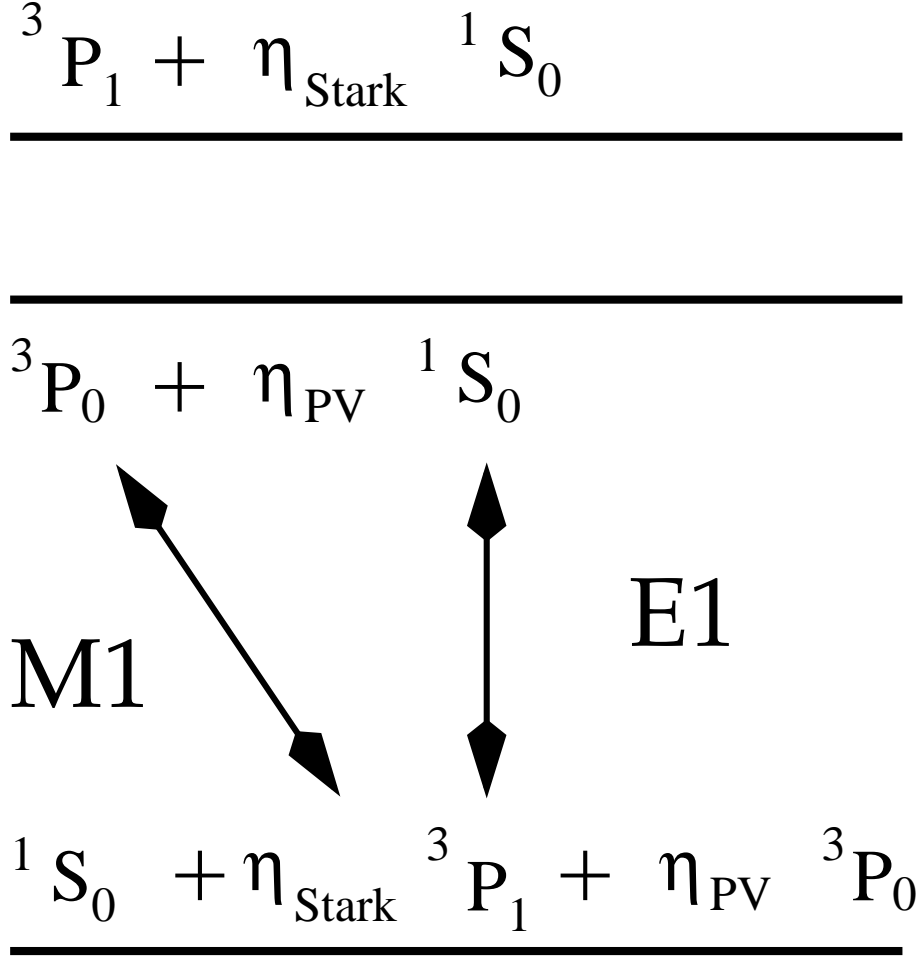,height=5in}
\caption{Stark- and parity violating admixtures to the first energy
         levels of beryllium-like uranium.}
\label{fig3}   
\end{figure}

In the last section we have seen that, by means of the Stark effect, we 
can induce a small parity conserving M1-transition between the
admixed $|^{\;3}P_1\rangle$ component of the ground state, 
and the $ |^{\;3}P_0\rangle$ state. In addition there is, though it is
much weaker, also a second transition which can be used, namely the 
pv admixed $|^{\;1}S_0\rangle$ component of the  
$ |^{\;3}P_0\rangle$ state to the Stark admixed 
$|^{\;3}P_1\rangle$ state within the ground state, see Fig.~\ref{fig3}. 
Suppose now that
we place our heavy ion in a Stark field $\vec E$ pointing into  
the z - direction, and 
orient the laser beam along the x-direction with polarization
vector $\epsilon = ( 0, \epsilon_y,\epsilon_z)$ (see Fig.~\ref{fig2}), where 
$\epsilon_y^2 + \epsilon_z^2 = 1$, then the transition rate from the
0+ ground state to the 0- first excited state is given by
\begin{eqnarray}
W_{0+ \rightarrow 0-} &=& W_{\rm laser;\; Stark} (M1) \epsilon_y^2
                         +W_{\rm laser;\; Stark+ pv} (E1) \epsilon_z^2
\end{eqnarray}
with 
\begin{eqnarray}
W_{\rm laser;\; Stark} (M1) &=& \frac{1}{2} \eta_{\rm Stark}^2
          \frac{ \pi c I_0}{\Gamma_{\rm laser} \omega_{1S0 \to 3P0}^2}
                | \langle ^3P_0 || M1 || ^3P_1 \rangle |^2
\nonumber \\
\nonumber \\
W_{\rm laser;\; Stark+ pv} (E1) &=& \frac{1}{2} 
          \eta_{\rm Stark}^2 \eta_{\rm pv}^2
          \frac{ \pi c I_0}{\Gamma_{\rm laser} \omega_{1S0 \to 3P0}^2}
              | \langle ^1S_0 || E1 || ^3P_1 \rangle |^2 \quad.
\end{eqnarray}
The remarkable point here is that the sort of transition being excited
depends fully on the polarization of the laser. If the
laser light is polarized along the z-direction 
($\epsilon_y=0$) only the parity violating
transition occurs.
However, the two amplitudes
differ by a factor $\eta_{\rm pv}^2$ which is of order $10^{-16}$, 
%
%
so the unwanted M1 transition has to be suppressed  by more than 
16 orders of magnitude by controlling the linear polarization 
of the laser, which is of course beyond the scope of 
present day technology.
Furthermore, as both are Stark induced, this is 
anyway an extremely weak transition. In Tab.~\ref{tab4} we refer 
again to $E_{\rm eff} = 10^{10}$ V/m and calculate the two induced 
transition rates. For uranium an intensity 
of at least $I = 10^{26}{\rm  W/cm}^2$ would be needed
for a counting
rate of 1000 photons per second and a detector of three meters
along the beam and a luminosity of $10^{11}$ ions per second
of the parity violating transition, which 
is of course completely impossible. We do think however, that the
basic mechanism is quite remarkable and we will search for cases for
which the Stark and parity violating admixture are larger. For
the lighter ions the situation is deteriorating for the
Stark induced E1 transition while it becomes slightly better in 
case of the M1 transition.

\section{Isolating parity violating effects via Stark induced
         excitations}
\begin{figure}
\psfig{figure=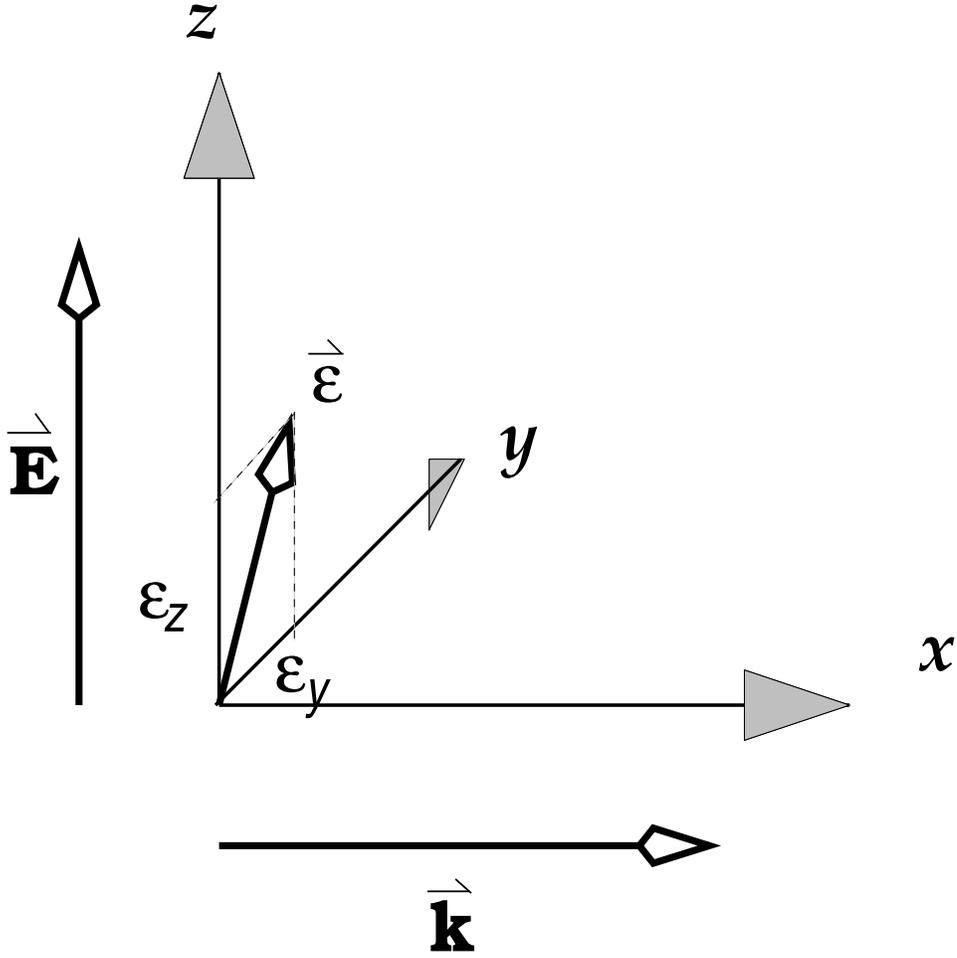,height=5in}
\caption{Geometry of the laser-induced experiments.}
\label{fig2}   
\end{figure}

In this chapter we combine parity violating effects 
with Stark quenching. The idea is the following:
We consider a beryllium-like heavy ion with zero nuclear spin 
and  a special laser
which emits two sorts of photons, one  with full energy and one
with half that energy. Such lasers are already in use.
The frequency splitting is due to special nonlinear crystals and 
the phase coherence is preserved during the process.
 The laser light is emitted in a
certain angle to the direction of the ion beam in order
to use the Doppler effect in such a way that the full energy is 
equal to the  $1s^22s2p^{\;3}P_0$ and the ${1s^22s^2} ^{\;1}S_0$
transition energy.
In order to keep a fixed phase between the Stark amplitude and
the parity violating amplitude, it is necessary to have a standing
wave of both sorts of photons with full and with half energy. 
The difficult
point is however that one needs a certain phase shift between the
two waves which comes from the fact that the interference part is
proportional to $\cos \phi$ with $\phi$ being the phase difference of the
two amplitudes. The phase shift needed can only be estimated when
the full two-photon transition is calculated. As we use only a
rough approximation we must leave this phase shift as a free parameter.
\newline
\newline
Furthermore the laser light must be linearly polarized. And the
spin of the two photons with half energy and the one with full
energy must be the same. The linear polarization has the big
advantage that it has components with negative as well as with positive
circular polarization and in this way there exist always two photons 
which  couple to  spin zero for the 2E1 transition.
\newline
\newline
Now the ion is placed in a Stark field to induce the admixture
of the $1s^22s2p^{\;3}P_1$ state to the $1s^22s^2{^{\;1}S_0}$ state.
The laser will coherently induce two transition amplitudes.
One is the M1 amplitude of the photons with full energy which is
Stark induced and the other is the parity violating amplitude which
is the 2E1 amplitude, from the two photons of half energy
\begin{eqnarray}
W &=& |  \sqrt{W_{\rm laser;\; Stark} (M1)} \epsilon_y 
      \; + \; \sqrt{W_{\rm laser:\; pv   } (2E1)} e^{i \phi}  |^2 
\nonumber \\
&\approx& W_{\rm laser;\; Stark} (M1) \epsilon_y ^2
\nonumber \\
 &&      +2 \epsilon_y \sqrt{W_{\rm laser;\; Stark} (M1)} 
           \sqrt{W_{\rm laser:\; pv   } (2E1)} \cos( \phi) \quad.
\end{eqnarray}
The coordinate system is shown in Fig \ref{fig2}: The direction
of the $\hat k$ vector of the photon is given by the x axis.
The Stark field points along the $z$-axis. 
The polarization vector $\vec \epsilon$  is then
lying in the $z$ - $y$ plane. The situation is most favorable if it points 
in the y-direction. We then may take simply the values
from Tab.~\ref{tab2}. We now use the maximum laser intensity
available, i.e., 
$I = 10^{17}{\rm  W/cm}^2$  with phase difference properly adjusted so that
we can set $\phi = 0$. Furthermore
we use a realistic Stark field of $E = 10^9$ V/m with a 
Doppler amplification of $10^3$ taken already into consideration. 
The width of the laser is taken to be $\Gamma_{\rm laser} = 1$ eV
in the moving frame of the ion.
We then get a counting
rate of 8000 photons per second. For a detector of three meters
along the beam and a luminosity of $10^{11}$ ions per second 
the parity violating contribution could be isolated
by reversing the sign of the linear dependent  E1-2E1 interference amplitude
\begin{equation}
W = (8.25703 \times 10^{+0} \pm  5.02553\times 10^{-02})
\; {\rm s}^{-1} \quad.
\end{equation}
This is an asymmetry of $6.1 \times 10^{-3}$.  For a counting
rate of 8000 photons per second and a detector of three meters
along the beam and a luminosity of $10^{11}$ ions per second, one would
need a run of $2\times 3.5 $ hours for a signal with a relative error of
1 \% and subsequently about 29 days for an error of 0.1 \%.
On the other hand, lets say it would be possible to construct
a laser with a line width of $\Gamma = 0.01 eV$ in the moving frame the
counting rate would be hundred times higher, thus reducing the time for
the 0.1\% - experiment to 7 hours. This shows how crucial all numbers given
here depend on the experimental conditions and on the atomic
structure. In both cases we can only present some plausible
estimates here. 
On the other hand, it is our aim to present a few concepts which
give some ideas how, with technical progress, such an
experiment might be feasible probably in another atomic system or in 
a heavy ion system with more than four or five  electrons. 
\newline
\newline
Note that the asymmetry will increase with every gain in laser power,
because the interference term increases as $I_0 ^{1.5}$, 
while the E1 amplitude increases only linearly with $I_0$. Furthermore 
reducing
the Stark field by one order of magnitude will amplify the
asymmetry also by an order of magnitude, but at the same time
the transition amplitude is reduced by two orders of magnitude too, so
one has to carefully optimize the parameter choice.
The asymmetry values
for other ions under the same conditions are displayed in
Tab.~\ref{tab5}. The asymmetry is decreasing for the lighter ions
down to iron and this means that these ions offer no favorable 
alternative to uranium. \newline \newline 
In summary we have examined a few theoretical concepts for 
parity violating experiments. We applied these concepts on beryllium-like 
uranium which unfortunately leads to presently unrealistic requirements on 
experimental device. Lighter ions than uranium are no alternative because the 
weak effects decrease without being compensated by an amplification by the
Stark quenching effect. 
But we stress again, that the discussed effect of the superposition of  one-
and  two-photon transition should be experimentally accessible in other more
favorable conditions where not necessarily heavy ions need to be involved, 
and this  will allow to study interesting phenomena.
\newline
\newline
\newline
The authors are thankful to I.~B.~Khriplovic, J.~Kluge,  D.~Liesen, 
and Th.~St\"ohlker
for useful discussions. This work was supported by BMBF.
%
\def \ajp#1#2#3{Am. J. Phys. {\bf#1}, #2 (#3)}
\def \apny#1#2#3{Ann. Phys. (N.Y.) {\bf#1}, #2 (#3)}
\def \app#1#2#3{Acta Phys. Polonica {\bf#1}, #2 (#3)}
\def \arnps#1#2#3{Ann. Rev. Nucl. Part. Sci. {\bf#1}, #2 (#3)}
\def \cmts#1#2#3{Comments on Nucl. Part. Phys. {\bf#1}, #2 (#3)}
\def \cn{Collaboration}
\def \cp89{{\it CP Violation,} edited by C. Jarlskog (World Scientific,
Singapore, 1989)}
\def \efi{Enrico Fermi Institute Report No. EFI}
\def \f79{{\it Proceedings of the 1979 International Symposium on Lepton and
Photon Interactions at High Energies,} Fermilab, August 23-29, 1979, ed. by
T. B. W. Kirk and H. D. I. Abarbanel (Fermi National Accelerator Laboratory,
Batavia, IL, 1979}
\def \hb87{{\it Proceeding of the 1987 International Symposium on Lepton and
Photon Interactions at High Energies,} Hamburg, 1987, ed. by W. Bartel
and R. R\\"uckl (Nucl. Phys. B, Proc. Suppl., vol. 3) (North-Holland,
Amsterdam, 1988)}
\def \ib{{\it ibid.}~}
\def \ibj#1#2#3{~{\bf#1}, #2 (#3)}
\def \ichep72{{\it Proceedings of the XVI International Conference on High
Energy Physics}, Chicago and Batavia, Illinois, Sept. 6 - 13, 1972,
edited by J. D. Jackson, A. Roberts, and R. Donaldson (Fermilab, Batavia,
IL, 1972)}
\def \ijmpa#1#2#3{Int. J. Mod. Phys. A {\bf#1}, #2 (#3)}
\def \ite{{\it et al.}}
\def \jdp#1#2#3{J.~Physique {\bf#1}, #2 (#3)}
\def \jpb#1#2#3{J.~Phys.~B~{\bf#1}, #2 (#3)}
\def \lkl87{{\it Selected Topics in Electroweak Interactions} (Proceedings of
the Second Lake Louise Institute on New Frontiers in Particle Physics, 15 -
21 February, 1987), edited by J. M. Cameron \ite~(World Scientific, Singapore,
1987)}
\def \ky85{{\it Proceedings of the International Symposium on Lepton and
Photon Interactions at High Energy,} Kyoto, Aug.~19-24, 1985, edited by M.
Konuma and K. Takahashi (Kyoto Univ., Kyoto, 1985)}
\def \mpla#1#2#3{Mod. Phys. Lett. A {\bf#1}, #2 (#3)}
\def \nc#1#2#3{Nuovo Cim. {\bf#1}, #2 (#3)}
\def \np#1#2#3{Nucl. Phys. {\bf#1}, #2 (#3)}
\def \pisma#1#2#3#4{Pis'ma Zh. Eksp. Teor. Fiz. {\bf#1}, #2 (#3) [JETP Lett.
{\bf#1}, #4 (#3)]}
\def \pl#1#2#3{Phys. Lett. {\bf#1}, #2 (#3)}
\def \pla#1#2#3{Phys. Lett. A {\bf#1}, #2 (#3)}
\def \plb#1#2#3{Phys. Lett. B {\bf#1}, #2 (#3)}
\def \pr#1#2#3{Phys. Rev. {\bf#1}, #2 (#3)}
\def \pra#1#2#3{Phys. Rev. A {\bf#1}, #2 (#3)}
\def \prc#1#2#3{Phys. Rev. C {\bf#1}, #2 (#3)}
\def \prd#1#2#3{Phys. Rev. D {\bf#1}, #2 (#3)}
\def \prl#1#2#3{Phys. Rev. Lett. {\bf#1}, #2 (#3)}
\def \prp#1#2#3{Phys. Rep. {\bf#1}, #2 (#3)}
\def \ptp#1#2#3{Prog. Theor. Phys. {\bf#1}, #2 (#3)}
\def \rmp#1#2#3{Rev. Mod. Phys. {\bf#1}, #2 (#3)}
\def \rp#1{~~~~~\ldots\ldots{\rm rp~}{#1}~~~~~}
\def \si90{25th International Conference on High Energy Physics, Singapore,
Aug. 2-8, 1990}
\def \slc87{{\it Proceedings of the Salt Lake City Meeting} (Division of
Particles and Fields, American Physical Society, Salt Lake City, Utah, 1987),
ed. by C. DeTar and J. S. Ball (World Scientific, Singapore, 1987)}
\def \slac89{{\it Proceedings of the XIVth International Symposium on
Lepton and Photon Interactions,} Stanford, California, 1989, edited by M.
Riordan (World Scientific, Singapore, 1990)}
\def \smass82{{\it Proceedings of the 1982 DPF Summer Study on Elementary
Particle Physics and Future Facilities}, Snowmass, Colorado, edited by R.
Donaldson, R. Gustafson, and F. Paige (World Scientific, Singapore, 1982)}
\def \smass90{{\it Research Directions for the Decade} (Proceedings of the
1990 Summer Study on High Energy Physics, June 25-July 13, Snowmass,
Colorado),
edited by E. L. Berger (World Scientific, Singapore, 1992)}
\def \tasi90{{\it Testing the Standard Model} (Proceedings of the 1990
Theoretical Advanced Study Institute in Elementary Particle Physics, Boulder,
Colorado, 3-27 June, 1990), edited by M. Cveti\v{c} and P. Langacker
(World Scientific, Singapore, 1991)}
\def \yaf#1#2#3#4{Yad. Fiz. {\bf#1}, #2 (#3) [Sov. J. Nucl. Phys. {\bf #1},
#4 (#3)]}
\def \zhetf#1#2#3#4#5#6{Zh. Eksp. Teor. Fiz. {\bf #1}, #2 (#3) [Sov. Phys. -
JETP {\bf #4}, #5 (#6)]}
\def \zpc#1#2#3{Zeit. Phys. C {\bf#1}, #2 (#3)}
\def \zpd#1#2#3{Zeit. Phys. D {\bf#1}, #2 (#3)}

\begin{table}
\begin{tabular}{|c|l|r||c|r|r|}
\hline
&&&&& \\
 $Z$  & Name & $A$ & $\eta_{\rm pv}$ & $E_0$[eV] & $E_1 [eV]$ \\
&&&&& \\
\hline
&&&&& \\
 26 & Fe &  56 & -9.92977 $\times 10^{-12}$ &  -22101.06 &  -22057.63 \\     
 36 & Kr &  84 & -5.38643 $\times 10^{-11}$ &  -43312.20 &  -43249.34 \\
 46 & Pd & 106 & -1.82287 $\times 10^{-10}$ &  -72101.15 &  -72016.56 \\
 56 & Ba & 138 & -6.02963 $\times 10^{-10}$ & -109047.58 & -108937.27 \\
 66 & Dy & 164 & -1.62324 $\times 10^{-09}$ & -154980.78 & -154839.01 \\ 
 76 & Os & 192 & -4.24362 $\times 10^{-09}$ & -211088.34 & -210907.49 \\ 
 82 & Pb & 208 & -7.41341 $\times 10^{-09}$ & -250322.86 & -250114.37 \\ 
 92 & U  & 238 & -1.91644 $\times 10^{-08}$ & -326604.06 & -326345.37 \\ 
\hline
\end{tabular}
\caption{
\label{tab1}
Parity mixing coefficients $\eta_{\rm pv}$ and energies for
the states $E_0= {1s^22s^2}^{\;1}S_0$ and $E_1=1s^22s2p^{\;3}P_0$
in stable beryllium-like heavy ions from Z=26 - Z=92 with zero
nuclear spin.}
\end{table}

\begin{table}
\begin{tabular}{|c|l|r||c|c|}
\hline
&&&& \\
 $Z$  & Name & $A$ 
 & $\frac{W_{\rm laser;\; nat}(E1M1)}{I_0^2}$ 
 &  $\frac{W_{\rm laser;\; pv} (2E1)}{I_0^2}$ \\
&&&& \\
\hline
&&&& \\
 26 & Fe &  56 & 6.10280$\times 10^{-26}$ &   8.80191$\times 10^{-44}$ \\
 36 & Kr &  84 & 2.25969$\times 10^{-25}$ &   7.71405$\times 10^{-42}$ \\
 46 & Pd & 106 & 2.07329$\times 10^{-25}$ &   6.92070$\times 10^{-41}$ \\
 56 & Ba & 138 & 1.07531$\times 10^{-25}$ &   3.35726$\times 10^{-40}$ \\
 66 & Dy & 164 & 4.48448$\times 10^{-26}$ &   8.90796$\times 10^{-40}$ \\
 76 & Os & 192 & 1.68055$\times 10^{-26}$ &   2.08910$\times 10^{-39}$ \\
 82 & Pb & 208 & 9.03356$\times 10^{-27}$ &   3.31807$\times 10^{-39}$ \\
 92 & U  & 238 & 3.24557$\times 10^{-27}$ &   7.64682$\times 10^{-39}$ \\
&&&& \\
\hline
\end{tabular}
\caption{
\label{tab2}
Approximated two-photon transition rates for the 
         unpolarized  transition into the $1s^22s2p^{\;3}P_0$
         state from the ground state via the E1M1 mode
         and the laser induced parity violating excitation
         via the 2E1 mode. All rates in the table are given in 
         $({\rm cm})^4 /(W^2 s)$.}
\end{table}

\begin{table}
\begin{tabular}{|c|l|r||r|c|c|}
\hline
&&&&& \\
 $Z$  & Name & $A$ & $\Delta E_0 [eV]$ 
      & $\Delta E_{ \rm Stark}$ [$eV m^2/V^2$ ] 
      & $\tau_{ \rm Stark}$ [$s V^2/m^2$] \\
&&&&& \\
\hline
&&&&& \\
 26 & Fe &  56 &  43.43 & 2.49577$\times 10^{-26}$ & 1.06087$\times 10^{+23}$\\
 36 & Kr &  84 &  62.86 & 3.66802$\times 10^{-26}$ & 1.13326$\times 10^{+22}$\\
 46 & Pd & 106 &  84.59 & 2.98420$\times 10^{-26}$ & 5.57576$\times 10^{+21}$\\
 56 & Ba & 138 & 110.31 & 1.94884$\times 10^{-26}$ & 4.95469$\times 10^{+21}$\\
 66 & Dy & 164 & 141.77 & 1.19901$\times 10^{-26}$ & 5.49249$\times 10^{+21}$\\
 76 & Os & 192 & 180.85 & 7.28323$\times 10^{-27}$ & 6.77734$\times 10^{+21}$\\
 82 & Pb & 208 & 208.49 & 5.39661$\times 10^{-27}$ & 7.98197$\times 10^{+21}$\\
 92 & U  & 238 & 258.69 & 3.31798$\times 10^{-27}$ & 1.10446$\times 10^{+22}$\\
&&&&& \\
\hline
\hline 
\end{tabular}
\caption{
\label{tab3}
Lifetime $\tau_{ \rm Stark}$ together with the values of the
energy separation $\Delta E_0$ of the unperturbed state and the 
Stark-perturbed state $\Delta E_{ \rm Stark}$.}
\end{table}

\begin{table}
\begin{tabular}{|c|l|r||c|c|}
\hline
&&&& \\
 $Z$  & Name & $A$ & $W(M1)/I_0$  $[(cm)^2 /(Ws)]$  
                   & $W(E1)/I_0$  $[(cm)^2 /(Ws)]$ \\
&&&& \\
\hline
&&&& \\
 26 & Fe &  56 &     1.81692$\times 10^{-13}$ &     1.31025$\times 10^{-31}$  \\ 
 36 & Kr &  84 &     5.60774$\times 10^{-13}$ &     9.57175$\times 10^{-30}$  \\ 
 46 & Pd & 106 &     4.67842$\times 10^{-13}$ &     7.80833$\times 10^{-29}$  \\ 
 56 & Ba & 138 &     2.37390$\times 10^{-13}$ &     3.70583$\times 10^{-28}$  \\ 
 66 & Dy & 164 &     1.00881$\times 10^{-13}$ &     1.00195$\times 10^{-27}$  \\ 
 76 & Os & 192 &     3.93855$\times 10^{-14}$ &     2.44801$\times 10^{-27}$  \\ 
 82 & Pb & 208 &     2.18246$\times 10^{-14}$ &     4.00671$\times 10^{-27}$  \\ 
 92 &  U & 238 &     8.25703$\times 10^{-15}$ &     9.72709$\times 10^{-27}$  \\ 
&&&& \\
\hline
\end{tabular}
\caption{
\label{tab4}
Transition coefficients for the Stark induced E1-transition
         W(E1) and the Stark induced M1-transition W(M1).}
\end{table}

\begin{table}
\begin{tabular}{|c|l|r|c|c|}
\hline
&&&& \\
 $Z$  & Name & $A$& $W$ [1/s] & Asymmetry = $\sqrt{\frac{W(2E1)}{W(M1)}}$ \\
&&&& \\
\hline
&&&& \\
26 & Fe &  56 & 1.81692$\times 10^{+02}$ $\pm$ 7.99809$\times 10^{-04}$ & 4.40201$\times 10^{-06}$ \\ 
36 & Kr &  84 & 5.60774$\times 10^{+02}$ $\pm$ 1.31542$\times 10^{-03}$ & 2.34572$\times 10^{-05}$ \\ 
46 & Pd & 106 & 4.67842$\times 10^{+02}$ $\pm$ 3.59877$\times 10^{-02}$ & 7.69229$\times 10^{-05}$ \\ 
56 & Ba & 138 & 2.37390$\times 10^{+02}$ $\pm$ 5.64617$\times 10^{-02}$ & 2.37843$\times 10^{-04}$ \\ 
66 & Dy & 164 & 1.00881$\times 10^{+02}$ $\pm$ 5.99549$\times 10^{-02}$ & 5.94311$\times 10^{-04}$ \\ 
76 & Os & 192 & 3.93855$\times 10^{+01}$ $\pm$ 5.73690$\times 10^{-02}$ & 1.45660$\times 10^{-03}$ \\ 
82 & Pb & 208 & 2.18246$\times 10^{+01}$ $\pm$ 5.38106$\times 10^{-02}$ & 2.46559$\times 10^{-03}$ \\ 
92 & Ur & 238 & 8.25703$\times 10^{+00}$ $\pm$ 5.02553$\times 10^{-02}$ & 6.08637$\times 10^{-03}$ \\ 
 &&&& \\
\hline
\end{tabular}
\caption{ 
\label{tab5}
Magnitude of the asymmetry which can be obtained
          in two Stark induced experiments where the sign
          of the Stark field is reversed in the second 
          experiment. The technical parameters are given
          in the text.}
\end{table}


\newpage
\begin{thebibliography}{99}

\bibitem{Al96} 
G. Altarelli,
{\em Status of precision tests of the standard model},
CERN-TH-96-265, Oct 1996. 32pp. \\
Talk given at the
NATO Advanced Study Institute on Techniques and 
Concepts of High-Energy Physics, St.
Croix, U.S. Virgin Islands, 10-23 Jul 1996 and \\
Talk given at Cracow International Symposium on
Radiative Corrections (CRAD 96), Cracow, Poland, 1-5 Aug 1996.\\ 
e-Print Archive: {\tt hep-ph/9611239}.


\bibitem{Ro95}
J.~L.~Rosner, \prd{53}{2724}{1996}. 


\bibitem{Le97}
B.~G.~Levi {\em 'Atomic parity experiment has its moment'}, Physics
Today, April 1997.


\bibitem{LNPBG96}
L.~N.~Labzowsky, A.~V.~Nefiodov, G.~Plunien, T.~Beier and G.~Soff,
\jpb{29}{3841}{1996}.

\bibitem{ASY95}
A.~N.~Artemyev, V.~M.~Shabaev, and V.~A.~Yerokin, \pra{52}{1884}{1995}.


\bibitem{BMPPGS97}
T.~Beier, P.~J.~Mohr, H.~Persson, G.~Plunien, M.~Greiner, G.~Soff,
\pla{236}{329}{1997}.

\bibitem{Schaefer}
A.~Sch\"afer and G.~Soff and P.~Indelicato and B.~M\"uller and W.~Greiner,
\pra{40}{7362}{1989}.

\bibitem{Maul}
M.~Maul, A.~Sch\"afer, W.~Greiner, and P.~Indelicato \pra{53}{3915}{1996}.

\bibitem{MPI93}
J.~P.~Marques, F.~Parente, and P.~Indelicato, \pra{47}{929}{1993}.

\bibitem{BB75}
M.~A.~Bouchiat and C.~Bouchiat, \jdp{36}{493}{1975}.

\bibitem{LW75}
R.~R.~Lewis and W.~L.~Williams, \plb{59}{70}{1975}.


\bibitem{BP83}
C.~Bouchiat and C.~A.~Piketty, \plb{128}{73}{1983}.


\bibitem{NMW88}
M.~C.~Noecker, B.~P.~Masterson, and C.~E.~Wieman, \prl{61}{310}{1988}.


\bibitem{MS83}
W.~J.~Marciano and A.~Sirlin, \prd{27}{552}{1983}.

\bibitem{Ro97}
J.~L.~Rosner\\
{\em New developments in precision electroweak physics}\\
 EFI-97-18, Apr 1997. 15pp. Submitted to Comments Nucl.Part.Phys.\\ 
e-Print Archive:{\tt  hep-ph/9704331}. 


\bibitem{De97}
A.~Deandrea, \plb{409}{277}{1997}. 
\end{thebibliography}
\end{document}